\documentclass[lettersize,journal]{IEEEtran}
\usepackage{amsmath,amsfonts}
\usepackage{algorithmic}
\usepackage{algorithm}
\usepackage{array}
\usepackage[caption=false,labelfont=sf]{subfig}
\usepackage{textcomp}
\usepackage{stfloats}
\usepackage{url}

\usepackage{verbatim}
\usepackage{graphicx}
\usepackage{cite}

\usepackage{arydshln}
\usepackage{multirow}
\usepackage{svg}
\usepackage{nccmath}
\usepackage{amssymb,amsmath,mathtools, amsthm}
\usepackage{booktabs}

\newcommand{\method}{PatchDSU}
\newcommand{\methodbase}{DSU}
\newcommand{\methodp}{$p$}

\newcommand{\vx}{\mathbf{x}}
\newcommand{\vmu}{\boldsymbol{\mu}}
\newcommand{\vsigma}{\boldsymbol{\sigma}}
\newcommand{\mSigma}{\boldsymbol{\Sigma}}
\newcommand{\vbeta}{\boldsymbol{\beta}}
\newcommand{\vgamma}{\boldsymbol{\gamma}}

\usepackage{tabularx}

\begin{document}

\title{\method: Uncertainty Modeling for Out of Distribution Generalization in Keyword Spotting}


\author{Bronya Roni Chernyak, Yael Segal, Yosi Shrem, Joseph Keshet
\thanks{All authors are with the Faculty of Electrical and Computer Engineering, Technion–Israel Institute of Technology, Israel}}
\markboth{Journal of \LaTeX\ Class Files,~Vol.~14, No.~8, August~2021}%
{Shell \MakeLowercase{\textit{et al.}}: A Sample Article Using IEEEtran.cls for IEEE Journals}


\maketitle

\begin{abstract}
Deep learning models excel at many tasks but rely on the assumption that training and test data follow the same distribution. This assumption often does not hold in real-world speech systems, where distribution shifts are common due to varying environments, recording conditions, and speaker diversity.

The method of Domain Shifts with Uncertainty (DSU) augments the input of each neural network layer based on the input feature statistics. It addresses the problem of out-of-domain generalization by assuming feature statistics follow a multivariate Gaussian distribution and substitutes the input with sampled features from this distribution. While effective for computer vision, applying DSU to speech presents challenges due to the nature of the data. Unlike static visual data, speech is a temporal signal commonly represented by a spectrogram - the change of frequency over time. This representation cannot be treated as a simple image, and the resulting sparsity can lead to skewed feature statistics when applied to the entire input.

To tackle out-of-distribution issues in keyword spotting, we propose PatchDSU, which extends DSU by splitting the input into patches and independently augmenting each patch. We evaluated PatchDSU and DSU alongside other methods on the Google Speech Commands, Librispeech, and TED-LIUM. Additionally, we evaluated performance under white Gaussian and MUSAN music noise conditions. We also explored out-of-domain generalization by analyzing model performance on datasets they were not trained on. Overall, in most cases, both PatchDSU and DSU outperform other methods. Notably, PatchDSU demonstrates more consistent improvements across the evaluated scenarios compared to other approaches.

\end{abstract}

\begin{IEEEkeywords}
Speech Recognition, Keyword Spotting, Out-Of-Distribution, Generalization, Domain Generalization
\end{IEEEkeywords}
\IEEEpubidadjcol
\section{Introduction}
Nowadays, many systems across various domains, such as vision and speech, leverage deep learning algorithms to achieve cutting-edge performance.  These algorithms rely significantly on the assumption that the distribution of the training data aligns with that of the testing data 
\cite{ben2010theory,vapnik1991principles}. Nevertheless, real-world applications often challenge this assumption. 

In the domain of speech, real-world data often diverges from the characteristics of the training set. Variations in speech rates, fundamental frequencies, tonalities, and dialects among speakers are common examples. Furthermore, external factors like microphone types or background noise not encountered during training can further alter the signal distribution. These discrepancies from the trained domain significantly impair the system's performance \cite{wilson2020survey} and call for directly addressing the model's robustness to unseen distributions. 


Keyword spotting (KWS), a core component in a wide variety of applications ranging from virtual assistants like Amazon's Alexa and Google Assistant, to smart home devices, is no stranger to these challenges. The goal of KWS is to detect a predefined set of keywords within a stream of user-pronounced utterances. 
In the literature, one line of research improves the performance of KWS in a general setup.  The work by Palaz \emph{et al.} \cite{palaz2016jointly} and Segal \emph{et al.} \cite{segal2019speechyolo} proposed to use Convolutional Neural Network (CNN) based approaches for both detection and localization of predefined keywords. Building on this, Fuchs \emph{et al.} \cite{fuchs2021cnn0} extended the work by Segal \emph{et al.} \cite{segal2019speechyolo} to unseen keywords. An alternative approach proposed by Vsvec \emph{et al.} \cite{vsvec2017relevance, vsvec2022spoken} involved using Recurrent Neural Network (RNN) solutions while incorporating the use of improved embeddings of searched terms. More recent work has advanced this direction by embedding keywords and integrating them with textual embeddings \cite{shin2022learning, nishu2023matching}. 
A more extensively studied setting focuses on improving the performance of KWS systems in small-footprint devices for detecting wake words that trigger the device's acoustic activation. 


Early work that introduced deep learning algorithms under the small-footprint constraint includes the work of Chen \emph{et al.}, Sainath \emph{et al.} \cite{chen2014small,sainath2015convolutional} and Zeng \emph{et al.} \cite{zeng2019effective}.
Later, Tang \emph{et al.} \cite{tang2018deep} proposed a successful approach that utilized residual deep-learning techniques and dilated convolutions. This approach was further built upon in subsequent work, such as Coucke \emph{et al.} \cite{coucke2019efficient}, who suggested a network based on dilated convolutions, gated activations, and residual connections. Later works, including TC-ResNet \cite{choi2019temporal}, TENet \cite{li2020small}, and BC-ResNet \cite{kim2021broadcasted}, proposed networks with residual connections with various temporal convolutions types. An important change introduced in the work of Tang \emph{et al.} \cite{tang2018deep} and subsequently adopted in other studies was to augment the input during training by performing time shifts and adding background noise to the signal whenever available in the dataset. While such augmentations can improve robustness to certain shifts, they primarily tackle specific aspects and may overlook the potential of leveraging statistical discrepancies in the signal for estimating potential distribution shifts.

With the daily use of technological devices integrated with KWS components, changes in the input characteristic are commonplace. Moreover, different signal interruptions, such as background noise or overlapping speech, affect the input to the system, leading to frequent distribution shifts. 

This works aims to improve existing capability of KWS systems, focusing on better domain generalization (DG) and out-of-domain generalization (OODG). 

Several learning paradigms are closely related to DG and OODG, with domain adaptation (DA) and zero-shot learning being key examples. Domain adaptation, which was studied in the context of different speech related tasks \cite{sun2017unsupervised, su2024corpus, wang2018unsupervised}, utilizes information from the source domain it is trained on, to improve system performance on a known target domain. In contrast, OODG tackles generalization to unknown target distributions, a more difficult scenario that closely resembles real-world application conditions. In contrast to OODG, zero-shot learning seeks to utilize knowledge from recognized keywords (or classes) to enhance the classification of unfamiliar keywords 
\cite{stafylakis2018zero, lee2023phonmatchnet,mazumder2021few}. In OODG, however, the target keywords are familiar, but the input distribution (such as recording setup and background noise) differs. Hence, OODG typically involves manipulating the training set features to extract robust representations that are less dependent on contextual information, which may be absent or different under distribution shifts. 



Several approaches were developed to address DG and OODG. The mix-up method\cite{zhang2017mixup} generates new samples by weighting combinations of pairs of samples within the batch. CrossGrad \cite{shankar2018generalizing} augments the input in the direction with the most domain change while maintaining minimal change to the label. BIN \cite{nam2018batch} proposed to replace batch normalization in the network by perturbing the input with instance-wise and batch-wise statistics. CSD \cite{piratla2020efficient}, introduced an architectural modification in the last layer to also learn a domain-specific component, which is discarded during inference, and a common component that generalizes to other distributions. More recently, Kim \emph{et al.} \cite{kim2022domain} proposed perturbing the input using batch frequency-wise statistics combined with layer normalization. Additionally, they introduced Freq-MixStyle, a variant of MixStyle \cite{zhou2021domain} that combines frequency-wise statistics. 

In the domain of vision, a promising approach to improve OODG was proposed in a recent study by Li \emph{et al.} \cite{li2022uncertainty}, demonstrating favorable performance compared to aforementioned methods such as Zhang \emph{et al.} \cite{zhang2017mixup} and Zhou \emph{et al.} \cite{zhou2021domain}. In their work, they propose their method DSU (Domain Shifts with Uncertainty), which can be treated as a neural network module that can be incorporated anywhere in the network. The core idea behind DSU is to leverage the variance of feature statistics to acquire out-of-domain relevant information, under the assumption that the original feature statistics conform to a multivariate Gaussian distribution. Accordingly, the method obtains and differentiates these statistics, facilitating the synthesis of novel features from the original distribution.


Contrary to images, which typically feature densely packed data, speech signals often exhibit sparser characteristics when subjected to visual pre-processing methods like spectrograms. In cases where the input domain exhibits such properties, sparser areas might obfuscate the broad information of the distribution. Instead, considering the distribution of multiple areas in the input can provide a better consideration of the shits. Motivated by this insight, we introduce {\method}, which extends the work of Li \emph{et al.} \cite{li2022uncertainty} by operating on the patch level rather than on the entire input. Our proposed modification begins by partitioning the input into patches. Following the original assumption of {\methodbase}, each patch conforms to its own multivariate Gaussian distribution. The method then proceeds by replacing the original input patches with new patches sampled from their corresponding distribution.

To evaluate the proposed method, we use the 12-keywords scenario of Google Speech Commands  v2 \cite{warden2018speech}. Furthermore, we created two additional datasets from Librispeech \cite{panayotov2015librispeech} and TED-LIUM\cite{hernandez2018ted}, conforming to the same keyword categories available in the target dataset, but with different unseen keywords. These efforts yield two additional unbalanced datasets that allow us to test (i) DG performance on each dataset with and without noise augmentation, and (ii) OODG performance on seen keywords. We evaluate {\method}, {\methodbase}, and compare them to Freq-MixStyle on ResNet-15.

This work extends previous studies in several aspects. Beyond adapting {\methodbase} to operate on patches, we analyze the robustness of both methods under noisy conditions, which was not addressed in \cite{li2022uncertainty}. Furthermore, while prior work on DG and OODG in speech typically relied solely on Google Speech Commands for both the seen and unseen distribution, we introduce two additional data sets that allow to evaluate performance on both zero-shot and out-of-distribution keywords.
Our implementation and datasets will be available online upon publication\footnote{\url{https://github.com/MLSpeech/PatchDSU}}.

\section{Background - DSU}
\label{background}

We begin with an overview of the method proposed by Li \emph{et al.} \cite{li2022uncertainty}, DSU (Domain Shifts with Uncertainty), to provide the necessary background for our extension, introduced in Section~\ref{method}. We assume that we work on a keyword spotting system that is implemented as a CNN, and that the input to each layer is called a \emph{feature map}. Typically, a feature map includes width, height, and several channels.

DSU posits that feature statistics, specifically their mean and standard deviation, are not static but can vary due to domain shifts. To improve the robustness of neural networks to potential variations (uncertainties), DSU models such variability in several aspects:
\begin{itemize}
    \item Probabilistic Modeling: Each feature statistic (mean and standard deviation) is modeled as a multivariate Gaussian distribution centered around its observed value, with variance estimated from mini-batch statistics. 
    \item Sampling During Training: During training, new feature statistics are sampled from these distributions, introducing variability that simulates potential domain shifts, as supported by previous work \cite{shen2021closed, wang2019implicit}.
\end{itemize}
This approach enables the model to learn feature representations that are resilient to statistical variations, thereby improving generalization to unseen domains. In practice, inputs can be replaced with sampled inputs drawn from the estimated distribution. Particularly, {\methodbase} can be treated as a module that can be applied between layers of deep neural networks.

Formally, denote by $\vx \in \mathbb{R}^{ C\times H \times W }$ the feature map, where $C$ represents the number of channels, $H$ is the height, and $W$ the width. In CNN training, we often work with min-batches of $B$ examples; therefore, we label each example in the mini-batch  $\vx_b$ for $1 \leq b \leq B$. We denote by $x_{b, c, h, w}$ the scalar value for specific values of $b, c, h$, and $w$. The channel-wise mean for each example in the batch is defined as: 
\begin{equation} 
\mu_{b,c} =\frac{1}{H W} \sum_{h=1}^{H} \sum_{w=1}^{W} x_{b, c, h, w}~,
\end{equation} 
where $\vmu_{b} \in \mathbb{R}^C$. Similarly, the standard deviation is defined as
\begin{equation}
\sigma^2_{b,c}  =\frac{1}{H W} \sum_{h=1}^{H} \sum_{w=1}^{W}\left(x_{b, c, h, w} - \mu_{b,c} \right)^2~,
\end{equation}
where $\vsigma^2_{b} \in \mathbb{R}^C$.

According to the DSU framework, each feature statistic is treated as a random variable drawn from a multivariate Gaussian distribution, with a mean vector that is distributed as $\mathcal{N}(\vmu_b, \mSigma^{\vmu})$, where $\mSigma^{\vmu}\in\mathbb{R}^{C \times C}$ is a diagonal matrix whose diagonal elements are the variances: 
\begin{equation}
\label{muvar}
\Sigma^{\vmu}_{c,c} =  
 \left( \mu_{b,c}- \frac{1}{B}\sum_{b=1}^{B} \mu_{b,c} \right)^2, ~~~ 1 \le c \le C~.
\end{equation}
Similarly, the deviation vector is distributed as $\mathcal{N}(\vsigma_b, \mSigma^{\vsigma})$ for $\mSigma^{\vsigma}\in\mathbb{R}^{C \times C}$ a diagonal matrix defined as
\begin{equation}
\label{sigvar}
\Sigma^{\vsigma}_{c,c} =  
 \left( \sigma_{b,c}- \frac{1}{B}\sum_{b=1}^{B} \sigma_{b,c} \right)^2, ~~~ 1 \le c \le C~.
\end{equation}
To allow a domain shift in the training phase, we shift each example in the mini-batch to have a mean $\vbeta_b\in\mathbb{R}^C$ sampled from Gaussian distribution $\vbeta_b\ \sim \mathcal{N}(\vmu_b, \mSigma^{\vmu})$ and a deviation $\vgamma_b\in\mathbb{R}^C$ sampled from $\vgamma_b\sim\mathcal{N}(\vsigma_b, \mSigma^{\vsigma})$.
Since the sampling operation is non-differentiable, the re-parameterization trick is incorporated \cite{kingma2013auto}. 
It is important to note that {\methodbase} is only used during training and applied with a probability $p\in[0,1]$. Particularly, since augmentation can affect in-domain generalization, $p$ is a tunable hyperparameter that controls the strength of the perturbation.  
Formally, {\methodbase} is define as
\begin{equation}
\text{DSU}_p(x_{b,c,h,w}) = 
\gamma_{b,c} \dfrac{x_{b,c,h,w} - \mu_{b,c}}{\sigma_{b,c}} + \beta_{b,c}.
\end{equation}
With probability  $p$, the {\methodbase} shift is applied to a given example in the mini-batch; otherwise, with probability $1 - p$, the example remains unaltered. The parameter  $p$ serves as a tunable hyperparameter of the model.

Lastly, {\methodbase} should not be confused with mini-batch Normalization (BN) \cite{ioffe2015batch}, or Instance-Normalization (IN) \cite{ulyanov2016instance}. In BN, the input is normalized using the mean and variance computed over the mini-batch for each feature channel, and then scaled and shifted using learnable parameters $\beta$ and $\gamma$. IN, on the other hand, normalizes inputs using instance-wise statistics. In contrast, while {\methodbase} also employs instance-wise normalization, it differs fundamentally by introducing stochasticity. Rather than using fixed parameters, it \emph{samples} $\beta$ and $\gamma$ from distributions estimated over mini-batch feature statistics. This mechanism enables {\methodbase} to model uncertainty and simulate potential domain shifts.
\section{Method - {\method}}
\label{method}
{\methodbase}\cite{li2022uncertainty} assumes that feature distribution follows a multivariate Gaussian distribution. However, when spectrogram-based pre-processing is applied in domains such as speech, the processed input can be sparse, leading to skewed feature statistics distributions.

\begin{figure*}[ht]
    \centering
\includegraphics[width=\linewidth,trim={0.35cm 0.1cm 0.41cm 0.4cm},clip]
{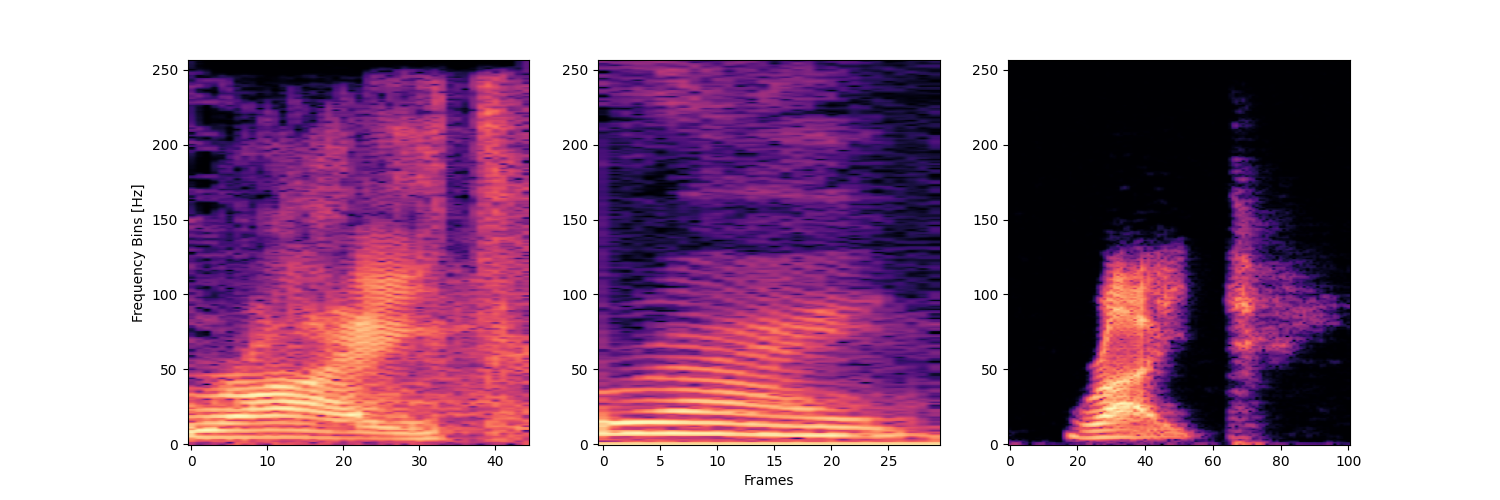}
\includegraphics[width=\linewidth,trim={0.35cm 0.1cm 0.41cm 1.6cm},clip]{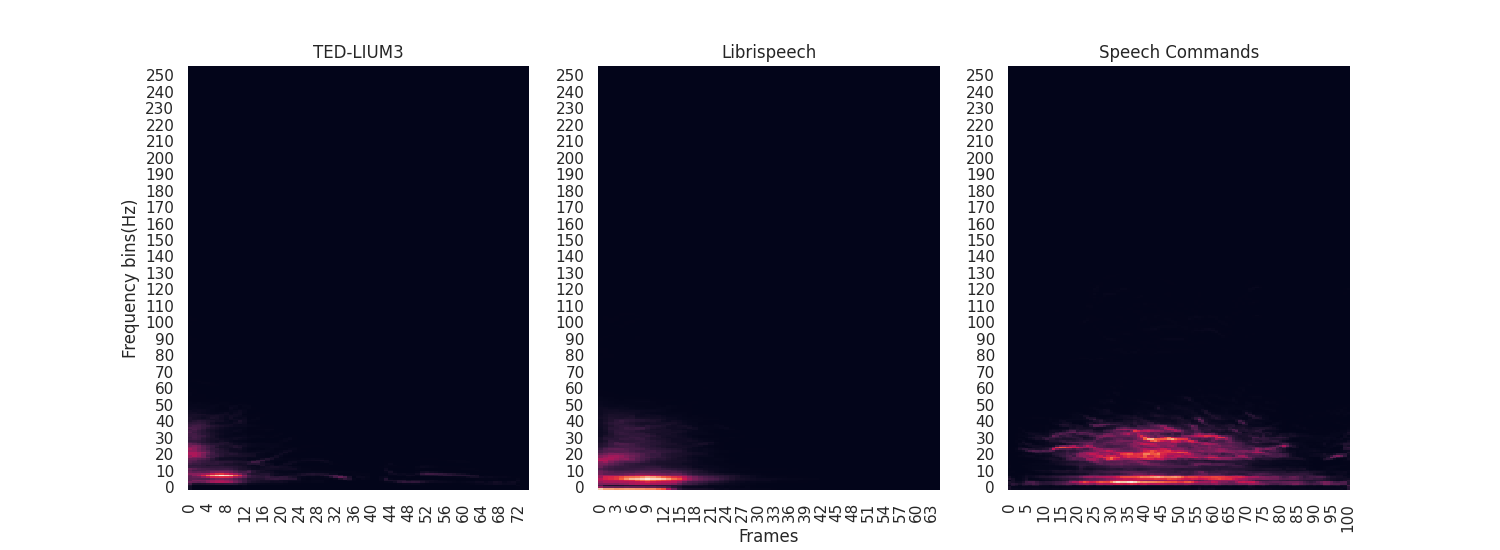}
\caption{Top: Log-Spectrogram of the keyword \emph{Right} of a single example from the test-set of TED-LIUM, Librispeech, and Speech Commands accordingly. Bottom: Mean log-spectrogram of the keyword \emph{Right}, computed over all test-set examples labeled \emph{Right} in TED-LIUM, LibriSpeech, and Speech Commands, respectively. 
In all plots, the y-axis represents frequency bins, the x-axis corresponds to time frames, and the pixel color encodes amplitude.}
\label{fig:spect}
\end{figure*}

This limitation is illustrated in Figure \ref{fig:spect}, where an analysis of the spectral characteristics of the keyword \emph{Right} across different datasets is presented. The top row displays a spectrogram of a single example from the keyword class in the test-set of each dataset. The bottom row presents a heatmap of the average spectrogram evaluated over the entire test set of each dataset. In both plots, the y-axis represents the frequency bins, and the x-axis depicts the time frames, with the pixel color indicating the amplitude of a bin at a specific time frame (lighter colors signifying higher amplitudes). 

An important insight from the figure is that the active frequency range, portrayed by the higher amplitude, is predominantly concentrated in the lower half of the spectrogram. Consequently, taking the feature statistics of the entire input will not accurately account for these spatial differences, which are essential for domain generalization.

Furthermore, there are substantial variations in the distribution of values across different datasets. Each dataset exhibits a distinct range of active frequencies, both within individual frames and throughout the entire time-frame of the test-set. Some datasets display a broad and diverse frequency range, while others are more constrained and concentrated. Moreover, the degree of overlap in the value ranges between datasets varies. While some pairs show minimal shared activity, others have greater alignment in their distributions.

These observations highlight potential challenges in out-of-domain generalization. In particular, generalizing between datasets with minimal overlap and disparate frequency ranges is likely to be more difficult than between those with more similar distributions. Consequently, relying solely on global feature statistics limits the ability to capture variations in local information, which can more efficiently portray the distributional characteristics difference between the mass and the tail.

Motivated by this, we propose extending {\methodbase} to the speech domain by applying it to localized regions of the spectrogram, which we refer to as \emph{patches}, forming the basis of our method, \emph{\method}. A patch denotes a small sub-region of the spectrogram that exhibits spatial stationarity. Particularly, this assumption is extended to patches, meaning that each patch feature statistics follow its multivariate Gaussian distribution. We hypothesize that sampling from the patch-level feature statistics, rather than the entire input distribution, will mitigate the effect of sparsity. Notably, this localized approach will lead to less skewed distributions, and would better capture local statistical properties, consequently providing more accurate uncertainty estimates. 
We now turn to formalize how {\method} works. 

Formally, denote by $\vx \in \mathbb{R} ^{C \times H \times W \times K}$ the feature map split into K patches. Specifically, the algorithm takes $x_{b, c, h, w}$ and splits it into $k$ patches. Each individual patch is denoted as $x_{k} \in \mathbb{R}^{B \times C  \times H_k \times W_k}$, such that $x_k$ has a height $H_k$ and width $W_k$. The dimensions are determined by the hyper-parameters $k_h, k_w \in\mathbb{Z}^+$ and are defined by:
\begin{equation}
H_k = \Big\lceil\frac{H}{k_h}\Big\rceil, W_k = \Big\lceil \frac{W}{k_w}\Big\rceil
\end{equation}

In the first layer, the input $x_{b, c, h, w}$ has undergone spectrogram-based preprocessing. Consequently, $W_k, H_k$ partition the input along the time and frequency dimensions, respectively.

We define the patch mean and standard deviation as follows:
\begin{equation} \label{new_mu}
\mu_{b,c,k} =\frac{1}{H_k W_k } \sum_{h=1}^{H_k} \sum_{w=1}^{W_k} x_{b, c, k, h, w}
\end{equation} 
and
\begin{equation} \label{new_sig}
\sigma^2_{b,c,k}  =\frac{1}{H_k W_k} \sum_{h=1}^{H_k} \sum_{w=1}^{W_k}\left(x_{b, c, k, h, w}-\mu_{b,c,k}\right)^2
\end{equation}
Where $\vmu_b\in \mathbb{R}^{C\times K}$ and $\vsigma^2_b\in \mathbb{R}^{C\times K}$.
Consequently, the diagonal matrices $\mSigma^{\vmu}$, $\mSigma^{\vsigma}$, whose elements are defined in Eq. \ref{muvar} and Eq. \ref{sigvar}, are constructed for each patch.


Finally, the {\method} is define as
\begin{equation}
\text{PatchDSU}(x_{b,c,k,h,w}) = 
\gamma_{b,c,k} \dfrac{x_{b,c,k,h,w} - \mu_{b,c,k}}{\sigma_{b,c,k}} + \beta_{b,c,k} ~,
\end{equation}
for every patch $k$. Similarly to {\methodbase} we applied this shift with probability $p$.

\section{Datasets}

Before turning to present the empirical results, we provide an overview of the datasets used in this study.
We begin by describing the Google Speech Commands V2  dataset, a well-known and widely used resource within the keyword spotting literature \cite{warden2018speech}. We then detail the creation process of two additional datasets derived from Librispeech \cite{panayotov2015librispeech}, and TED-LIUM \cite{hernandez2018ted}. Specifically, speech segments containing the same keywords as those found in the 12-class Speech Commands dataset were extracted. The \emph{Silences} class and any classes with fewer than 20 examples were excluded from these datasets.
It is important to note that the \emph{Unknown} class contains utterances different from those in the Google Speech Commands dataset. Still, it remains consistent between the Librispeech and TED-LIUM datasets. These additional datasets enable a broader range of experiments by incorporating more diverse speech data, allowing for a more comprehensive evaluation of out-of-distribution performance. All datasets have a sampling rate of 16 kHz.

\subsection{Google Speech Commands V2}
Google speech commands V2 \cite{warden2018speech} is composed of 105k utterances of one-second duration split into 35 classes. The dataset exhibits variability in recording quality, ranging from noisy to clearer recordings. In our analysis, we focus on the 12-classes scenario, where only ten specific keyword classes are used:  \emph{Down}, \emph{Up}, \emph{Left}, \emph{Right}, \emph{On}, \emph{Off}, \emph{Yes}, \emph{No}, \emph{Go}, and \emph{Stop}. The additional two classes are \emph{Unknown}, which includes the remaining 25 classes, and \emph{Silences}, which contains background noise from the data.

\subsection{Librispeech - 11 classes}
\label{sub:libri}
Librispeech \cite{panayotov2015librispeech} is a corpus containing 1000 hours of read-speech and corresponding transcriptions. In this paper, we train the model with the \emph{train-100} and \emph{train-360} sets, containing 100 and 360 hours of clean speech. For test, we used the \emph{train-other-500} set containing 500 hours of speech with variability in speaker accents, recording quality, and background noise than the former sets. To provide an alternative distribution of the same classes as in Google Speech Commands, we aligned the audio files using Montreal Forced Aligner\footnote{\url{https://github.com/MontrealCorpusTools/Montreal-Forced-Aligner}}. Then, we extract words corresponding to all the classes except \emph{Silences}. Finally, we filter files containing keywords longer than 1 second to adhere to the experimental setup of Speech Commands and the training procedure of previous work. Contrary to the Speech Commands dataset, the classes are not balanced. The Librispeech - 11 classes dataset contains approximately eight hours of speech recordings in the training set, two hours in the validation set, and ten hours in the test set.

\subsection{TED-LIUM - 8 classes}
TED-LIUM \cite{hernandez2018ted} is a corpus comprising 2351 TED talks, which amounts to 452 hours of speech. In contrast to the other datasets, TED-LIUM provides an alternative distribution containing conversational speech rather than read speech and may vary in the recording quality between samples. The corpus includes automatic alignments created using the Kaldi toolkit. However, the toolkit was not able to annotate many segments, and the alignments were not at the word level. Since there are no fully annotated options available, we overcome this by using wav2vec 2.0 \cite{baevski2020wav2vec} to transcribe the files. Then, we use both MMS \cite{pratap2023scaling} and Wav2Vev2 over the generated transcriptions to align the files. Finally, we follow the process described in Section \ref{sub:libri} to extract the classes, also resulting in an unbalanced dataset. Upon examining the number of available examples for validation and test, we removed from the dataset three classes that had less than 20 examples:  \emph{Left}, \emph{Yes}, \emph{Stop}. The TED-LIUM - 8 classes dataset consists of 9.4 hours for training, approximately 1.3 hours for the validation set, and 1.3 hours for the test set.

\section{Experiments}
\label{sec:exp}

In this section, we compare {\method},{\methodbase}, Freq-MixStyle \cite{kim2022domain} across different scenarios. For the backbone model of the experiments, we used ResNet-15 (as detailed in the ``Implementation details" in subsection A) due to its prominent performance on the KWS task using the Google Speech Commands dataset, as well as the availability of a detailed, publicly available implementation. Accordingly, we use ResNet-15 trained without any OODG methods as our baseline.

For the implementation of $\method$ we selected two patch sizes: $k_w=3, k_h=7$ and $k_w=10, k_h=6$. Their selection was guided by validation performance, as both consistently outperformed alternative configurations across multiple runs. Importantly, these patch sizes were chosen to maintain a structured partitioning of the frequency axis into multiple bands, enabling more effective modeling of frequency-specific patterns while varying the degree of temporal segmentation.


Furthermore, since no implementation was available for the work of Kim et al. \cite{kim2022domain}, we implemented their approach, which combines instance-wise frequency normalization with MixStyle, herein referred to as \emph{Freq-MixStyle}. The hyperparameters were set to $\lambda=0.5$ and application probability of $ p=0.8$, in accordance with the original work.

Throughout the experimental section, the average F1-score is reported for each method. This evaluation is conducted on the clean data (the original test set) and data augmented with various background noise types at different signal-to-noise ratios (SNR). Additionally, our experiments include two noise settings: White Gaussian Noise (WGN) and music noises from the MUSAN library \cite{snyder2015musan}.

The remainder of this section continues with an assessment of in-corpus generalization. Next, model performance is analyzed when the test set is augmented with various types of noise. Following that, we conduct an ablation study illustrating the effect of different {\methodp} values on the performance of {\method} and {\methodbase} across all datasets. In the last subsection, out-of-domain generalization (OODG) is examined by evaluating how models trained on one dataset perform on the test sets of the other datasets.

\subsection{Implementation details}
 Our training setup follows the one proposed by \cite{tang2018deep}. Specifically, we used their ResNet-15 model, a residual network \cite{he2015deep} composed of 13 layers, feature maps of size 45, and dilation with an exponential sizing schedule. Each layer consists of $3\times3$ convolution followed by ReLU activation \cite{agarap2018deep} and batch-normalization \cite{ioffe2015batch}. Six of these layers incorporate residual blocks. For models that were trained with {\method}, {\methodbase}, or Freq-MixStyle, we applied the method's operation on the input before the convolution in each layer. The network input initially goes through a forty-dimensional Mel-Frequency Cepstrum Coefficient (MFCC) with an FFT size of 512, a window size of 480, and a stride of 160. 
 
Training on the Google Speech Commands dataset was conducted with Google’s pre-processing guidelines. Background noise samples provided with the dataset were randomly added to each training instance with a probability of 0.8 per epoch. Additionally, a random time shift within the range of 
$[-100,100]$ milliseconds was applied. For the remaining datasets, background noises were not available; To simulate similar conditions, each recording was trimmed by an additional $K$ milliseconds from both ends. This approach ensures that when a time shift is applied, the resulting sample is introduced with an additional utterance, part of an utterance, or noise. Furthermore, to avoid excessive trimming of the keyword itself, $K$ was set to be the minimal value between 0.05 seconds and half of the signal length.

Consequently, no time shift was applied to the Google Speech Commands dataset during inference. In contrast, for the other datasets, where time shifts may introduce additional utterances, we report results with and without time shifts to assess performance under more challenging and realistic conditions. 

Lastly, models were trained for 300 epochs on LibriSpeech with early stopping, and for 200 epochs on the other datasets. We employed the Stochastic Gradient Descent (SGD) optimizer with a momentum of 0.9, weight decay of 0.001, and a batch size of 100. Also, we used a learning rate schedule combining warm-up and cosine annealing, where the learning rate increased from zero to 0.1 \cite{loshchilov2016sgdr,goyal2017accurate}.

\subsection{In corpus performance}

In this section, we evaluate the generalization of different methods across three datasets: Google Speech Commands, TED-LIUM, and Librispeech. In addition to reporting the F1 score for all datasets, we also report, for Librispeech and TED-LIUM, the average F1 score across five runs with different random seeds under time-shifted input conditions. Time-shifting introduces potential noise or part of another utterance, which may impact the model's performance. Results are shown in the Table \ref{tab:cleanacc} (where time augmentation is abbreviated as Time Agt.). 

When evaluating the performance on the Google Speech Commands dataset, the F1-scores of all models are comparable. However, on Librispeech, both {\method} models performed best. Notably, {\method}$(k_h\!=\!6, k_w\!=\!10)$ model showcased the highest results, improving the performance on the clean test-set by 0.7\% over {\methodbase}, a 1.29\% over baseline, and 1\% over Freq-MixStyle. Furthermore, {\methodbase} also improves generalization compared to the baseline and Freq-MixStyle. 

On the Libriseech time-shifted test-set, all models displayed lower performance, validating that it posed more of a challenge than the standard test-set. Additionally, the gap between {\method}$(k_h\!=\!6, k_w\!=\!10)$ and the other models is even more pronounced, with a 0.95\% improvement from {\methodbase} and over a 1\% improvement from the baseline model and Freq-MixStyle.

In contrast, results on the TED-LIUM dataset were higher with time-shifting than without, potentially indicating consistent noise levels in this dataset compared to Librispeech. Still, all models performed better than the baseline model, except Freq-MixStyle, which showed a decline of 1.72\% and 1.23\% on the clean and time-shifted test sets, respectively. This suggests that utilizing the deterministic feature on low-resource data may negatively impact generalization performance. In addition, on the clean test set, both {\method}$(k_h\!=\!6, k_w\!=\!10)$ and {\methodbase} models outperformed other methods by over 1\%, and were comparable to each other. Concurrently, {\method}$(k_h\!=\!3, k_w\!=\!7)$ performed comparably to the baseline model. 

Nevertheless, on the time-shifted test-set, {\method}$(k_h\!=\!6, k_w\!=\!10)$ outperformed other methods. Demonstrating improved performance over {\method} by 0.8\% and by above 3\% over Freq-MixStyle and the baseline.

\subsection{Robustness to noise}
Next, we evaluate the performance of the methods when the test set samples are augmented with White Gaussian Noise (WGN) and music noise from the MUSAN dataset. Specifically, the augmentation is applied using signal-to-noise ratios (SNRs) ranging from 20 dB to -5 dB. The lowest SNR in this range is selected based on when a sufficient degradation in F1 score is observed—i.e., the point at which further reduction in SNR would not provide additional insight into performance deterioration. Results are shown in Table \ref{tab:gcomwgn}-Table \ref{tab:tedmusan}. It is important to note that the models were not trained with these specific noises. However, the Google Speech Commands dataset includes background noise during training, unlike the LibriSpeech and TED-LIUM datasets. In contrast, the time-shifts applied to LibriSpeech and TED-LIUM may introduce potential noise or parts of other utterances.

Turning to the results on Google Speech Commands, surprisingly, while at higher SNRs under WGN, Freq-MixStyle performed better than other approaches. Specifically, at SNR of -5 dB and MUSAN noises, it consistently underperformed compared to all other methods, trailing behind the baseline model by over 3\%. In contrast, the {\methodbase} model performed similarly to the {\method} models at higher SNRs under both noise conditions. However, their performance gap in lower SNRs increased to approximately 1\%, with both approaches significantly outperforming Freq-MixStyle and the baseline, particularly in the presence of MUSAN noises.


When focusing on other datasets, the performance under noise considerably deteriorates. For Librispeech, when augmented with WGN, all models except {\methodbase} exhibit lower performance than the baseline model at SNRs below 10 dB, deviating by over 5\% when the SNR is 5 dB. The only exception is the time-shift test set, with an SNR of 10 dB. This behavior persists for {\method}$(k_h\!=\!6, k_w\!=\!10)$ and Freq-MixStyle when evaluated on TED-LIUM with WGN augmentations. Nevertheless, on Librispeech with MUSAN noises, the {\method}$(k_h\!=\!6, k_w\!=\!10)$ model exceeded the performance of other models, except for lower SNR scenarios where the {\methodbase} model had comparable results.

On TED-LIUM, results varied, and there was a higher standard deviation among runs. In some cases, {\methodbase} had the best performance, and in others, it was one of the {\method} models. Particularly, {\method}$(k_h\!=\!7, k_w\!=\!3)$ was predominantly better than {\method}$(k_h\!=\!6, k_w\!=\!10)$, which became more prominent in lower SNRs. Notably, when examining the impact of MUSAN noise types on both Librispeech and TED-LIUM, it appears to pose less of a challenge than the WGN setting.


Overall, except for Google Speech Commands under WGN and Librispeech under MUSAN noises, {\methodbase} had a favorable or comparable performance to other models under different noise conditions on SNRs below 20 dB. However, under SNR of 20 dB and on Librispeech under MUSAN noises, {\method}$(k_h\!=\!6, k_w\!=\!10)$ model had favorable or comparable performance.

\begin{table*}[ht]
  \caption{Performance of several methods on Google Speech Command (denoted as Speech Commands)  12 classes, Librispeech 11 classes, and TED-LIUM 8 classes (in the brackets after each name we added the number of classes). Performance under time shifts is also reported for the latter two datasets and denoted as Time Agt. 
  }
  \label{tab:cleanacc}
  \centering
  \begin{tabular}{@{\extracolsep{-6pt}}lcccccccc}
  \toprule
    \multirow{2}{*}{Method}& \multicolumn{2}{c}{Speech Commands(12)} & \multicolumn{3}{c}{Librispeech(11)} & \multicolumn{3}{c}{TED-LIUM(8)}\\
    &$p$&F1&$p$&F1&Time Agt. F1.&$p$&F1&Time Agt. F1.\\
    
  \hline
  {\method}$(k_h\!=\!6, k_w\!=\!10)$&0.4&\textbf{98.43}&0.2&\textbf{95.91}&\textbf{94.65$\pm$0.07}&0.1&\textbf{76.33}&\textbf{82.07$\pm$0.40}\\{\method}$(k_h\!=\!7, k_w\!=\!3)$&0.3&98.30&1.0&95.79&94.10$\pm$0.05&0.3&74.97&80.44$\pm$0.40\\{\methodbase}&0.5&98.42&1.0&95.19&93.70$\pm$0.00&0.4&76.31&81.27$\pm$0.03\\
    Freq-MixStyle&0.8&98.32&0.8&94.88&93.52$\pm$0.02&0.8&73.08&77.57$\pm$0.48\\
  ResNet-15&-&98.09&-&94.62&93.56$\pm$0.03&-&74.62&78.80$\pm$0.10\\      
  
    \bottomrule
  \end{tabular}
\end{table*}

  \begin{table*}[ht]
  \caption{Noise performance evaluated on the Google Speech Commands (denoted as Speech Commands) dataset (12 classes). F1 scores are reported on the test set under WGN noise with various SNRs.}
  \label{tab:gcomwgn}
  \centering
  \begin{tabular}{@{\extracolsep{-6pt}}lcccccc}
  
  \toprule
     Method&$p$&20 dB& 10 dB&5 dB&0 dB&-5 dB\\
     \hline
PatchDSU$(k_h\!=\!6, k_w\!=\!10)$&0.4&96.31$\pm$0.08&95.34$\pm$0.13&94.21$\pm$0.19&92.67$\pm$0.12&90.11$\pm$0.48\\
PatchDSU$(k_h\!=\!7, k_w\!=\!3)$&0.3&96.38$\pm$0.07&95.17$\pm$0.3&93.95$\pm$0.32&92.44$\pm$0.20&90.36$\pm$0.37\\
DSU&0.5&96.59$\pm$0.19&95.45$\pm$0.15&94.55$\pm$0.20&\textbf{93.24$\pm$0.17}&\textbf{91.20$\pm$0.40}\\
Freq-MixStyle&0.8&\textbf{97.34$\pm$0.10}&\textbf{96.17$\pm$0.06}&\textbf{94.64$\pm$0.10}&92.11$\pm$0.37&85.63$\pm$0.40\\
ResNet-15&-&96.23$\pm$0.22&94.84$\pm$0.26&93.51$\pm$0.28&91.42$\pm$0.11&88.46$\pm$0.27\\
    \bottomrule
  \end{tabular}
\end{table*}

  \begin{table*}[ht]
  \caption{Noise performance evaluated on the Google Speech Commands (denoted as Speech Commands) dataset (12 classes). F1 scores are reported on the test set under MUSAN noise with various SNRs.
  }
  \label{tab:gcommusan}
  \centering
  \begin{tabular}{@{\extracolsep{-6pt}}lcccccc}
  
  \toprule
Method&$p$&20 dB&10 dB&5 dB&0 dB&-5 dB\\
\hline
PatchDSU$(k_h\!=\!6, k_w\!=\!10)$&0.4&96.38$\pm$0.17&94.36$\pm$0.23&92.12$\pm$0.19&89.39$\pm$0.40&85.27$\pm$0.51\\
PatchDSU$(k_h\!=\!7, k_w\!=\!3)$&0.3&96.06$\pm$0.09&94.35$\pm$0.28&92.32$\pm$0.15&89.65$\pm$0.18&85.94$\pm$0.38\\
DSU&0.5&\textbf{96.53$\pm$0.19}&\textbf{94.52$\pm$0.21}&\textbf{92.83$\pm$0.25}&\textbf{90.02$\pm$0.55}&\textbf{86.64$\pm$0.51}\\
Freq-MixStyle&0.8&87.77$\pm$0.23&85.93$\pm$0.25&84.07$\pm$0.39&79.01$\pm$0.57&70.64$\pm$0.65\\
ResNet-15&-&95.94$\pm$0.10&93.03$\pm$0.20&90.49$\pm$0.43&87.24$\pm$0.35&82.57$\pm$0.63\\
    \bottomrule
  \end{tabular}
\end{table*}

  \begin{table*}[ht]
  \caption{Performance of the methods under noise conditions is evaluated on the Librispeech (11 classes) dataset. F1 scores on the test set are reported under WGN noise at various SNRs. Additionally, F1 scores are presented for the noisy signal with time shifts (denoted with ``Agt.'' near the SNR value), and without time shifts (denoted only with the SNR value).}
  \label{tab:libriwgn}
  \centering
  \tabcolsep=0.16cm
  \begin{tabular}{@{\extracolsep{-6pt}}lccccccc}
  
  \toprule
 Method&$p$& 20 dB Agt.&20 dB& 10 dB Agt.&10 dB&5 dB Agt.&5 dB\\
 \hline
PatchDSU$(k_h\!=\!6, k_w\!=\!10)$&0.2&85.22$\pm$0.04&87.40$\pm$0.03&66.9$\pm$0.16&68.13$\pm$0.05&53.41$\pm$0.14&53.81$\pm$0.16\\
PatchDSU$(k_h\!=\!7, k_w\!=\!3)$ &1.0&84.02$\pm$0.16&86.36$\pm$0.19&64.14$\pm$0.21&64.49$\pm$0.06&51.98$\pm$0.14&50.26$\pm$0.17\\
DSU&1.0&\textbf{85.39$\pm$0.10}&\textbf{87.54$\pm$0.07}&\textbf{73.07$\pm$0.05}&\textbf{74.41$\pm$0.01}&\textbf{60.40$\pm$0.01}&\textbf{60.34$\pm$0.11}\\
Freq-MixStyle&0.8&84.01$\pm$0.1&86.19$\pm$0.14&66.17$\pm$0.13&67.05$\pm$0.10&52.40$\pm$0.16&51.14$\pm$0.09\\
ResNet-15&-&82.56$\pm$0.05&83.26$\pm$0.00&67.60$\pm$0.16&66.51$\pm$0.04&59.31$\pm$0.13&56.74$\pm$0.03\\
    \bottomrule
  \end{tabular}
\end{table*}

  \begin{table*}[ht]
  \caption{Performance of the methods under noise conditions is evaluated on the Librispeech (11 classes) dataset. F1 scores on the test set are reported under MUSAN noise at various SNRs. Additionally, F1 scores are presented for the noisy signal with time shifts (denoted with ``Agt.'' near the SNR value) and without time shifts (denoted only with the SNR value).}
  \label{tab:librimusan}
  \centering
  \tabcolsep=0.16cm
  \begin{tabular}{@{\extracolsep{-6pt}}lccccccccc}
  \toprule
  Method&$p$& 20 dB Agt.&20 dB& 10 dB Agt.&10 dB&5 dB Agt.&5 dB&0 dB Agt.&0 dB\\
  \hline

PatchDSU$(k_h\!=\!6, k_w\!=\!10)$&0.2&\textbf{91.36$\pm$0.00}&\textbf{92.84$\pm$0.02}&\textbf{83.52$\pm$0.03}&\textbf{85.10$\pm$0.11}&\textbf{76.14$\pm$0.06}&\textbf{77.55$\pm$0.12}&\textbf{65.92$\pm$0.05}&\textbf{66.84$\pm$0.28}\\
PatchDSU$(k_h\!=\!7, k_w\!=\!3)$ &1.0&90.69$\pm$0.07&92.39$\pm$0.04&82.18$\pm$0.05&83.85$\pm$0.12&74.44$\pm$0.03&76.17$\pm$0.05&64.53$\pm$0.14&65.38$\pm$0.2\\
DSU&1.0&90.45$\pm$0.01&92.15$\pm$0.02&82.71$\pm$0.00&84.55$\pm$0.06&75.43$\pm$0.01&77.49$\pm$0.15&65.08$\pm$0.16&66.53$\pm$0.05\\
Freq-MixStyle&0.8&89.54$\pm$0.06&90.86$\pm$0.10&80.45$\pm$0.20&81.94$\pm$0.13&72.12$\pm$0.27&73.84$\pm$0.30&61.59$\pm$0.17&62.46$\pm$0.13\\
ResNet-15&-&90.13$\pm$0.00&90.83$\pm$0.01&80.97$\pm$0.20&81.25$\pm$0.12&73.59$\pm$0.11&74.00$\pm$0.23&63.98$\pm$0.03&63.92$\pm$0.11\\
    \bottomrule
  \end{tabular}
\end{table*}

  
  

  \begin{table*}[ht]
  \caption{Performance of the methods under noise conditions is evaluated on the TED-LIUM (8 classes) dataset. F1 scores on the test set are reported under WGN noise at various SNRs. Additionally, F1 scores are presented for the noisy signal with time shifts (denoted with ``Agt.'' near the SNR value) and without time shifts (denoted only with the SNR value).}
  \label{tab:tedwgn}
  \centering
  \tabcolsep=0.16cm
  \begin{tabular}{@{\extracolsep{-6pt}}lcccccccccc}
  
  \toprule
     Method&$p$& 20 dB Agt.&20 dB& 10 dB Agt.&10 dB&5 dB Agt.&5 dB&0 dB Agt.&0 dB
     &\\
     \hline

PatchDSU$(k_h\!=\!6, k_w\!=\!10)$&0.1&\textbf{75.13$\pm$0.86}&64.79$\pm$0.15&64.29$\pm$2.34&53.08$\pm$0.75&56.95$\pm$1.14&47.78$\pm$1.38&47.4$\pm$1.31&40.99$\pm$1.62&\\
PatchDSU$(k_h\!=\!7, k_w\!=\!3)$&0.3&73.97$\pm$0.71&\textbf{67.31$\pm$0.74}&65.82$\pm$0.43&59.23$\pm$0.82&58.94$\pm$0.91&\textbf{54.57$\pm$0.90}&49.1$\pm$0.79&\textbf{45.83$\pm$0.33}&\\
DSU&0.4&71.63$\pm$1.13&63.58$\pm$0.29&\textbf{66.55$\pm$1.01}&\textbf{59.48$\pm$0.28}&\textbf{60.91$\pm$0.42}&53.22$\pm$0.79&\textbf{50.85$\pm$1.13}&45.44$\pm$1.24&\\
     Freq-MixStyle&0.8&71.79$\pm$0.54&66.98$\pm$0.49&63.62$\pm$0.78&58.26$\pm$0.99&55.26$\pm$1.8&49.47$\pm$1.13&42.24$\pm$1.97&37.79$\pm$0.56&\\
     ResNet-15&-&58.91$\pm$1.05&62.55$\pm$0.55&51.08$\pm$1.0&53.24$\pm$0.4&49.53$\pm$0.6&49.44$\pm$0.9&45.06$\pm$0.9&43.87$\pm$0.97&\\
    \bottomrule
  \end{tabular}
\end{table*}

  \begin{table*}[ht]
  \caption{Performance of the methods under noise conditions is evaluated on the TED-LIUM (8 classes) dataset. F1 scores on the test set are reported under MUSAN noise at various SNRs. Additionally, F1 scores are presented for the noisy signal with time shifts (denoted with ``Agt.'' near the SNR value) and without time shifts (denoted only with the SNR value).}
  \label{tab:tedmusan}
  \centering
  \tabcolsep=0.16cm
  \begin{tabular}{@{\extracolsep{-6pt}}lcccccccccccc}
  
  \toprule
     Method&$p$& 20 dB Agt.&20 dB& 10 dB Agt.&10 dB&5 dB Agt.&5 dB&0 dB Agt.&0 dB&-5 dB Agt.&-5 dB\\
     \hline

PatchDSU$(k_h\!=\!6, k_w\!=\!10)$&0.1&\textbf{79.19$\pm$0.82}&71.93$\pm$1.33&72.65$\pm$1.14&65.2$\pm$1.06&67.67$\pm$1.32&59.73$\pm$2.14&59.65$\pm$1.19&52.74$\pm$1.12&50.96$\pm$2.42&46.49$\pm$1.49&\\
PatchDSU$(k_h\!=\!7, k_w\!=\!3)$&0.3&78.47$\pm$0.59&\textbf{72.88$\pm$0.76}&73.68$\pm$1.05&68.26$\pm$1.24&69.46$\pm$1.27&62.72$\pm$1.69&62.18$\pm$1.67&\textbf{58.20$\pm$1.37}&52.96$\pm$1.29&49.42$\pm$1.08\\
DSU&0.4&78.43$\pm$0.67&71.35$\pm$0.87&\textbf{74.02$\pm$1.71}&\textbf{69.44$\pm$0.44}&\textbf{71.15$\pm$0.86}&\textbf{64.37$\pm$1.62}&\textbf{63.94$\pm$0.62}&58.16$\pm$1.68&\textbf{54.64$\pm$1.81}&\textbf{50.0$\pm$1.89}&\\
     Freq-MixStyle&0.8&75.46$\pm$0.88&70.95$\pm$0.96&71.6$\pm$0.75&67.06$\pm$0.86&66.87$\pm$1.02&62.06$\pm$1.27&59.79$\pm$0.88&55.77$\pm$0.31&53.09$\pm$1.93&47.57$\pm$0.68\\
     ResNet-15&-&63.60$\pm$0.69&65.8$\pm$0.73&58.35$\pm$1.05&61.45$\pm$0.98&53.89$\pm$1.3&55.85$\pm$1.16&47.63$\pm$0.58&49.29$\pm$0.76&41.58$\pm$1.59&43.12$\pm$1.47\\
    \bottomrule
  \end{tabular}
\end{table*}

\subsection{Impact of application probability on generalization}
\begin{figure*}[htb]
  \centering
  \raisebox{0\height}{}
  \subfloat[Google Speech Commands
  \label{fig:gcom_p_abl_clean}]{\includegraphics[width=0.36\textwidth]{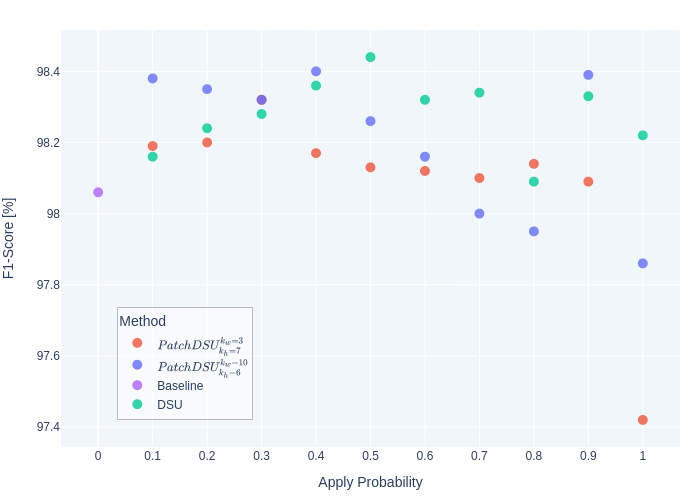}}
  \subfloat[Librispeech \label{fig:libri_p_abl_clean}]{\includegraphics[width=0.36\textwidth]{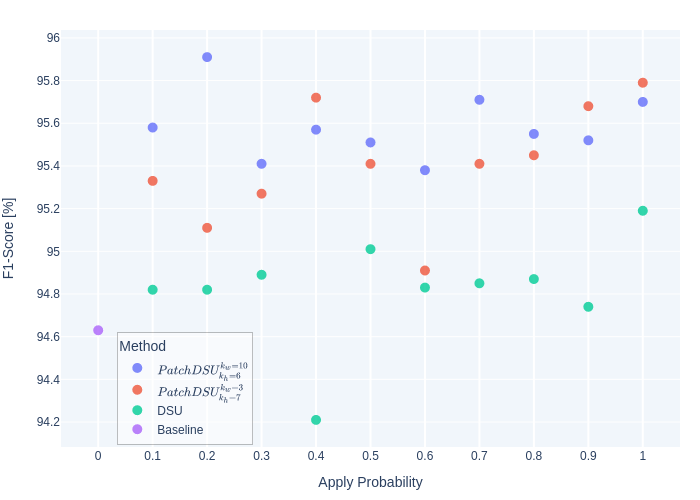}}
    \subfloat[Librispeech Agt.\label{fig:libri_p_abl_aug}]{
    \includegraphics[width=0.36\textwidth]{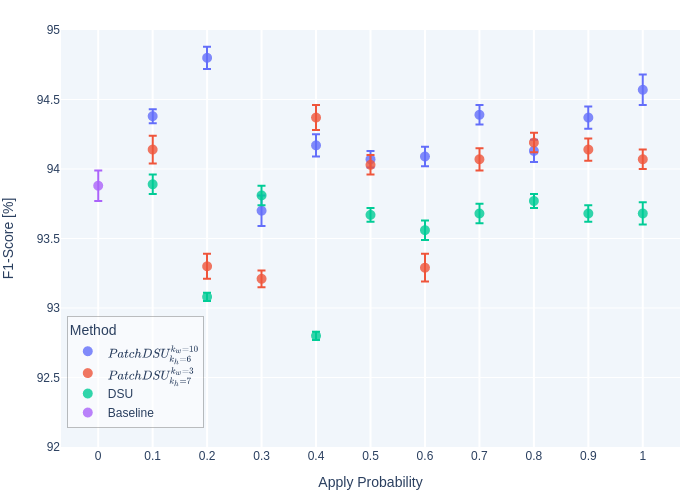}
    }
    \par
    \subfloat[TED-LIUM\label{fig:ted_p_abl_clean}]{
    \includegraphics[width=0.4\textwidth]{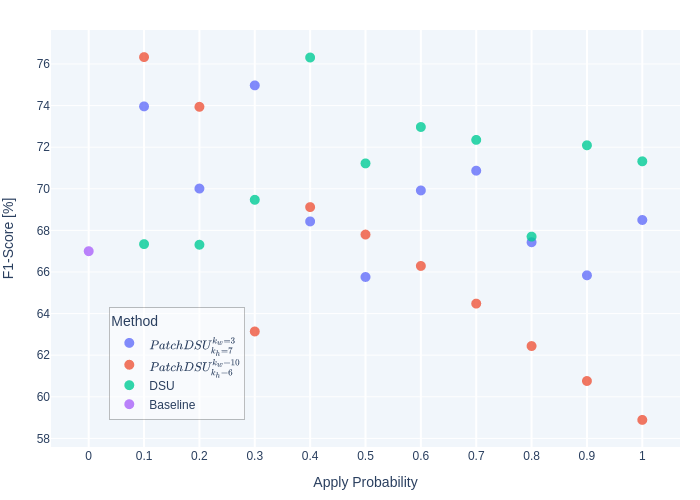}
    }
    \subfloat[TED-LIUM Agt.\label{fig:ted_p_abl_aug}]{
    \includegraphics[width=0.4\textwidth]{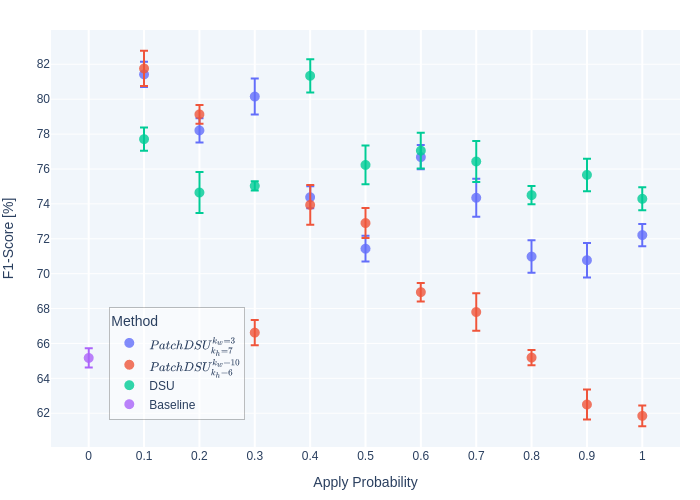}
    }\par
    \caption{Effect of application probability $p$ on each dataset. The y-axis represent the F1-score as a function of the application probability of the model.}
    \label{fig:ablation_p}

\end{figure*}

We now turn to analyze the impact of varying the application probability $p$ (i.e., the likelihood of augmenting the input using the method, during training) on model performance for {\method} and {\methodbase} across three datasets. These findings are illustrated in Figure \ref{fig:ablation_p}. As shown in Figure \ref{fig:gcom_p_abl_clean}, the baseline performance on the Google Speech Commands dataset is relatively high compared to the other two datasets, as reflected in its initial F1-score. Consequently, most models showed similar performance, with the majority of $p$-values improving the baseline. However, some {\method} models showed performance degradation at high application probabilities, likely due to overly aggressive augmentation, which negatively impacted generalization. Notably, models trained with {\method}$(k_h\!=\!6, k_w\!=\!10)$ consistently outperformed those trained with {\method}$(k_h\!=\!7, k_w\!=\!3)$. Particularly, models with augmentation probabilities between 0.1 and 0.4 had better performance compared to other value choices, with the exception of 0.9. 

For Librispech's clean-test-set, in Figure. \ref{fig:libri_p_abl_clean}, {\method} models generally outperformed {\methodbase}. Additionally, while {\method}$(k_h\!=\!7, k_w\!=\!3)$ models improved with higher $p$-values, other methods showed minimal variation. This was also consistent for the test-set with time-shifts. Furthermore, in the time-shift scenario, as shown in Figure \ref{fig:libri_p_abl_aug}, {\methodbase} consistently underperformed (except for $p=0.1$), whereas training with {\method} enhanced the generalization, except for several models trained with {\method}$(k_h\!=\!7, k_w\!=\!3)$. 

When analyzing the performance on the TED-LIUM dataset in Figures \ref{fig:ted_p_abl_clean} and \ref{fig:ted_p_abl_aug}, there is a broader range of F1-scores compared to other datasets, showcasing a larger variation among models. Models trained with {\method}$(k_h\!=\!7, k_w\!=\!3)$ demonstrated a consistent negative linear trend in F1-scores across different $p$-values on both the clean and time-shifted test-sets. In contrast, {\method}$(k_h\!=\!6, k_w\!=\!10)$ models maintained better performance with lower $p$ values. Similarly, models trained with {\methodbase} generally performed best with higher $p$ values on the clean-test set, but lower values (0.1 and 0.4) were more effective on the time-shifted test-set.

Overall, as shown in Figure 2,   {\method}$(k_h\!=\!6, k_w\!=\!10)$ model consistently demonstrates improved generalization across the values of 0.1 and 0.2. Conversely, the other two training methods do not exhibit such behavior. Nevertheless, calibration should rely on the validation set.

\subsection{Out of domain generalization}
We conclude the results section by evaluating the models' out-of-domain generalization capabilities. Our objective is to assess the performance of models on unseen domains without additional training. For these scenarios, we analyze two aspects: (i) their performance relative to models trained on the unseen domain and (ii) how the models fare against a baseline ResNet-15 model when evaluated on the unseen test set.

To assess cross-domain generalization, each model trained on a specific dataset is evaluated against the other two datasets. 
Recall that the \emph{Unknown} class varies across datasets. Google Speech Commands contains a closed set of 25 fixed keywords, whereas the other datasets include a larger and more diverse set of keywords with varying occurrences. Consequently, performance is analyzed both through standard evaluation and on keywords, defined as the subset of the test set excluding the \emph{Unknown} class.

The following metrics are considered: the F1-score of the entire test set (denoted as F1), and the F1-score evaluated only on keywords (denoted as F1$_{keys}$ in the tables, and referred to as \emph{keywords} or \emph{keywords test set} in the text), and the F1-score on keywords when the dataset includes time shifts (applied to TED-LIUM and LibriSpeech, denoted as \emph{Time Agt.}).

\begin{figure}
    \centering

  \raisebox{0\height}{}
  \subfloat[Performance of GSC models on Librispeech
  \label{fig:gcom_ood_on_libri}]{\includegraphics[width=\linewidth]{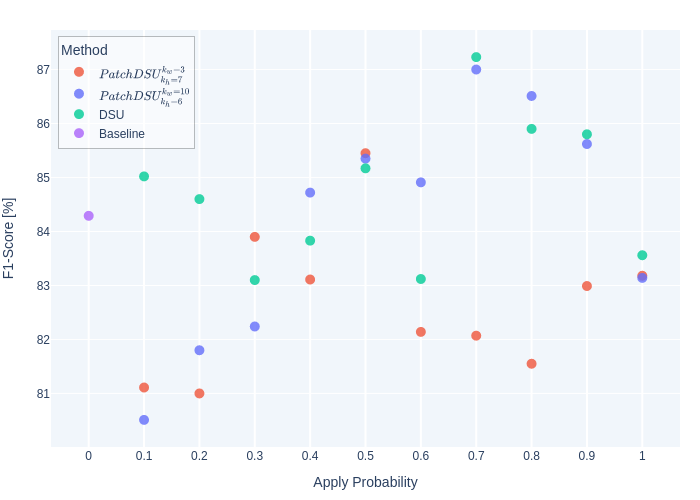}}
    
  \subfloat[Performance of GSC models on TED-LIUM
  \label{fig:gcom_ood_on_ted}]{\includegraphics[width=\linewidth]{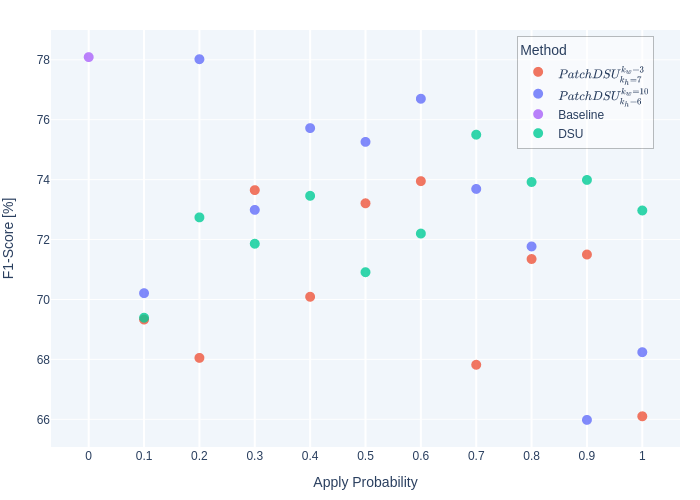}}\par
    \caption{Performance of models trained on Google Speech Commands (GSC) and tested on Librispeech and TED-LIUM keywords test splits (test-set without the Unknown class).}
    \label{fig:gcom_ood}
\end{figure}

Since each dataset has a different number of keywords (classes), we evaluate only the keywords available in each dataset. The detailed results are presented in Table \ref{tab:cross}. As the evaluation is conducted on models from Table \ref{tab:cleanacc}, which exhibited the best in domain generalization, we also include additional top-performing models below a dashed line in some evaluations to highlight those that demonstrated strong OODG performance. We emphasize that their performance was also the best on the out-of-domain validation set. Recall that their performance on the dataset they were trained on is available in Figure \ref{fig:ablation_p}. 
Lastly, we also provide the performance on the keywords test set of all models trained with {\method} and {\methodbase} on the counterpart datasets in Figures \ref{fig:gcom_ood}-\ref{fig:ted_ood}.

  \begin{table*}[ht!]
  \caption{Cross domain generalization performance. We report the F1-score performance of models that were trained on one dataset (column \emph{Trained On}) and evaluated on the other datasets (column \emph{Tested On}). The performance is evaluated on the clean test-set, the clean test-set without the unknown class (denoted as \emph{Clean F1$_{keys}$}), under WGN with SNR of -5, and MUSAN music noises with -5 of SNR. Google Speech Commands is denoted as Speech Commands}
  \label{tab:cross}
  \centering
  \begin{tabular}{@{\extracolsep{-6pt}}llclccccccc}
  
  \toprule 
 Trained On&Tested On&$\#Classes$&Method&$p$&F1& F1$_{keys}$&Time Agt. F1$_{keys}$\\
\hline

  Librispeech&TED-LIUM&8&{\method}$(k_h\!=\!6, k_w\!=\!10)$&0.2&62.28&88.07&78.16$\pm$0.01\\
 && &{\method}$(k_h\!=\!7, k_w\!=\!3)$&1.0&\textbf{66.61}&\textbf{89.53}&77.03$\pm$0.01\\
 && &{\methodbase}&1.0&62.74&84.63&\textbf{78.45$\pm$0.01}\\
 && &Freq-MixStyle&0.8&59.85&83.84&74.59$\pm$0.01 \\
 && &ResNet-15&-&65.48&87.66&73.98$\pm$0.01\\
 
 \midrule
      Librispeech&Speech Commands&11&{\method}$(k_h\!=\!6, k_w\!=\!10)$&0.2&79.41&82.55&-\\
 && &{\method}$(k_h\!=\!7, k_w\!=\!3)$&1.0&78.24&81.91&-\\ 
 && &{\methodbase}&1.0&80.23&83.33&-\\
 && &Freq-MixStyle&0.8&\textbf{80.61}&\textbf{84.06}&-\\
 && &ResNet-15&-&76.90&80.12&-\\
\hdashline
 &&&{\method}$(k_h\!=\!6, k_w\!=\!10)$&0.9&80.61&83.59\\

 \midrule
 Speech Commands&Librispeech&11&{\method}$(k_h\!=\!6, k_w\!=\!10)$&0.4&\textbf{80.76}&84.72&\textbf{81.03$\pm$0.001}\\
 && &{\method}$(k_h\!=\!7, k_w\!=\!3)$&0.3&79.28&83.91&79.55$\pm$0.001\\  
 && &{\methodbase}&0.5&80.58&\textbf{85.17}&80.42$\pm$0.001\\&&&Freq-MixStyle&0.8&75.97&80.82&76.52$\pm$0.001\\
 && &ResNet-15&-&73.92&84.29&79.66$\pm$0.001\\

 \midrule
  Speech Commands&TED-LIUM&8& {\method}$(k_h\!=\!6, k_w\!=\!10)$&0.4&44.30&75.72&62.44$\pm$0.02\\&& &{\method}$(k_h\!=\!7, k_w\!=\!3)$&0.3&41.19&73.65&61.20$\pm$0.01\\ && &{\methodbase}&0.5&39.38&70.91&62.25$\pm$0.01\\
 && &Freq-MixStyle&0.8&40.44&67.65&57.89$\pm$0.01\\
 && &ResNet-15&-&\textbf{51.69}&\textbf{78.09}&\textbf{63.46$\pm$0.03}\\\hdashline
 
 &&&{\method}$(k_h\!=\!6, k_w\!=\!10)$&0.2&47.32&78.02&63.06$\pm$0.01\\
 
 \midrule
TED-LIUM&Librispeech&8& {\method}$(k_h\!=\!6, k_w\!=\!10)$&0.1&\textbf{78.39}&\textbf{81.78}&\textbf{78.89$\pm$0.007}\\
 && &{\method}$(k_h\!=\!7, k_w\!=\!3)$&0.3&73.42&76.38&72.81$\pm$0.001\\  
 && &{\methodbase}&0.4&69.25&72.07&70.36$\pm$0.001\\
 && &Freq-MixStyle&0.8&68.40&71.97&70.53$\pm$0.001\\
 && &ResNet-15&-&73.03&76.20&74.44$\pm$0.001\\    
 \midrule
 TED-LIUM&Speech Commands&8& {\method}$(k_h\!=\!6, k_w\!=\!10)$&0.1&32.54&33.39&-\\
 && &{\method}$(k_h\!=\!7, k_w\!=\!3)$&0.3&41.49&43.60&-\\  
 && &{\methodbase}&0.4&20.38&19.82&-
 \\ && &Freq-MixStyle&0.8&\textbf{70.77}&\textbf{75.02}&-\\
 && &ResNet-15&-&48.03&50.99&-\\
\hdashline
&&&{\method}$(k_h\!=\!6, k_w\!=\!10)$&0.7&66.80&70.80\\

    \bottomrule
  \end{tabular}
\end{table*}

\subsubsection{Models trained on Librispeech}
We first assess the performance of models trained on Librispeech. Notably, the Unknown class distribution of Librispeech is closer to that of Google Speech Commands than to TED-LIUM. This is reflected in the baseline F1 performance (ResNet-15 in the table), which is higher on the Google Speech Commands dataset than on TED-LIUM. Additionally, the gap between the F1-score on the full test set and the F1-score on the keywords test set (excluding the \emph{Unknown} class) exceeds 30\% for the TED-LIUM dataset.
%
In terms of performance on TED-LIUM, {\method}$(k_h\!=\!7, k_w\!=\!3)$ model outperformed other approaches and the baseline by over 4\% and 1\%, respectively, on the whole test-set and the keywords split. However, the {\methodbase} model was better by 1.5\% on the augmented split. When comparing {\method}$(k_h\!=\!6, k_w\!=\!10)$ model to the {\methodbase} model, while they had comparable performance on the time-shift setting and the full test-set, {\method}$(k_h\!=\!6, k_w\!=\!10)$ model had higher performance on the keywords subset. Furthermore, the Freq-MixStyle model had lower performance than the baseline model on the full test-set and the keyword test-set.
Interestingly, all models surpassed the best TED-LIUM-trained model, which had a Clean F1$_{keys}$ (keyword F1-score) of 76.34\%.

\begin{figure}
    \centering

  \raisebox{0\height}{}
  \subfloat[Performance of Librispeech models on GSC
  \label{fig:libri_ood_on_gcom}]{\includegraphics[width=\linewidth]{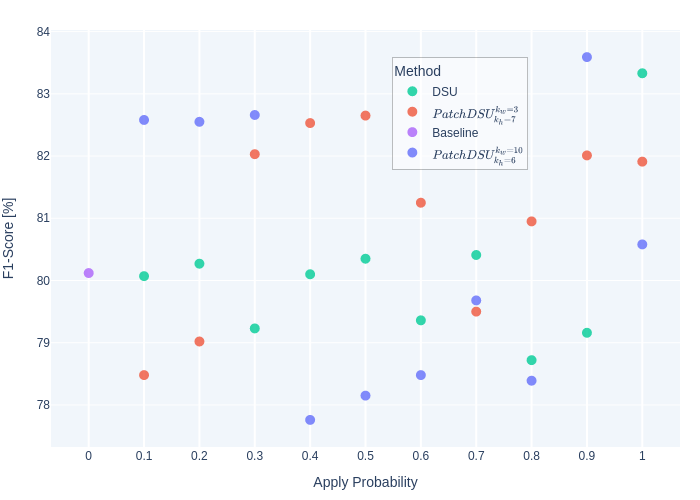}}
    
  \subfloat[Performance of Librispeech models on TED-LIUM
  \label{fig:libri_ood_on_ted}]{\includegraphics[width=\linewidth]{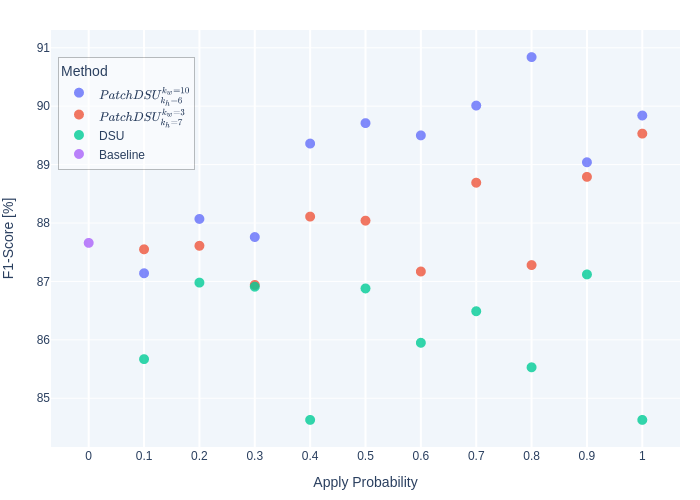}}\par
    \caption{Performance of models trained on Librispeech and tested on GSC and TED-LIUM keywords test splits (test-set without the Unknown class).}
    \label{fig:libri_ood}
\end{figure}

When evaluating results on the Google Speech Commands dataset, the Freq-MixStyle model performed best, outperforming {\method} models by up to 2\% and {\methodbase} model by 0.7\% on the keywords test-set. When inspecting the top performing model in Figure \ref{fig:libri_ood_on_gcom}, {\method}$(k_h\!=\!6, k_w\!=\!10)$ with $ p=0.9$, it had comparable performance on the test-set. However, it scored 0.4\% less on the keywords split than the Freq-MixStyle model. Despite this, both {\method}$(k_h\!=\!6, k_w\!=\!10)$ models exhibited better generalization on the Librispeech dataset than the Freq-MixStyle model by 2\%. This pattern highlights a trade-off between OODG and generalization.


\subsubsection{Models trained on Google Speech Commands} 
When inspecting performance on Librispeech, we can see that {\method}$(k_h\!=\!6, k_w\!=\!10)$ and {\methodbase} models performed best. Notably, improving full test-set performance by 4\% and 5\% over Freq-MixStyle and baseline, respectively, achieving gains of  4\% and 1\% on the keyword split, and enhancing performance by 5\% and 1.5\% in the augmented setting. While they demonstrated comparable performance on the overall test-set, {\methodbase} performed better on the keywords subset, and {\method}$(k_h\!=\!6, k_w\!=\!10)$ was better in the time-shift setting. Additionally, the model trained with Freq-MixStyle underperformed by nearly 4\% relative to the baseline model.

Turning to examine results on the TED-LIUM dataset, consistent with previous observations, the Unknown class continued to impact overall model performance negatively, having a gap of over 20\% between the test-set and the keywords test-set. This is due to the distribution of the class being very different between the datasets, both in terms of recording quality and in the types of keywords within the ``Unknown'' class. Furthermore, unexpectedly, the baseline model performed the best among the methods across all scenarios. In light of this, we inspected the top performing model among {\method} and {\methodbase} on this test-set, {\method}$(k_h\!=\!6, k_w\!=\!10)$ with $p=0.2$ (added under the dashed line) showed comparable results, whereas {\method}$(k_h\!=\!6, k_w\!=\!10)$ with $p=0.5$ exhibited a 2\% lower performance on the keyword test-set by 2\% while improving the time-shift test-set by 3\%.

\subsubsection{Models trained on TED-LIUM}
When assessing the performance on the Librispeech test set, we can see that overall, the {\method}$(k_h\!=\!6, k_w\!=\!10)$ model surpassed other methods, achieving over 5\% improvement across all settings. 

Surprisingly, the next best performance, which was comparable to that of the {\method}$(k_h\!=\!7, k_w\!=\!3)$ model, was achieved by the baseline model. Both models demonstrated close results, except on the time-shift test, where the baseline model performed better.

\begin{figure}
    \centering

  \raisebox{0\height}{}
  \subfloat[Performance of TED-LIUM models on GSC
  \label{fig:ted_ood_on_gcom}]{\includegraphics[width=\linewidth]{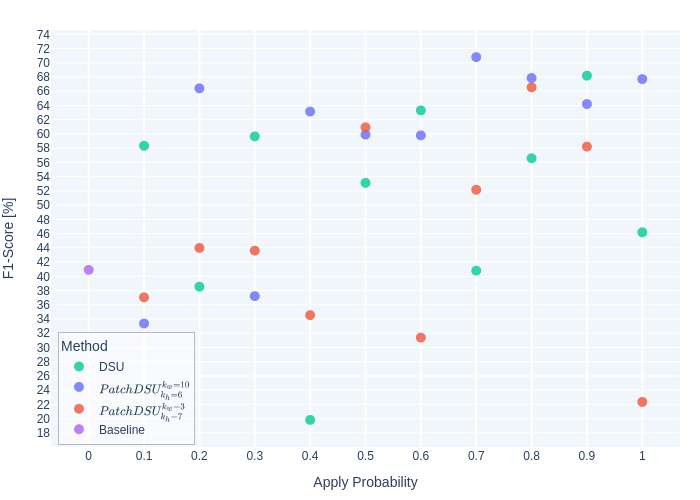}}
    
  \subfloat[Performance of TED-LIUM models on Librispeech
  \label{fig:ted_ood_on_libri}]{\includegraphics[width=\linewidth]{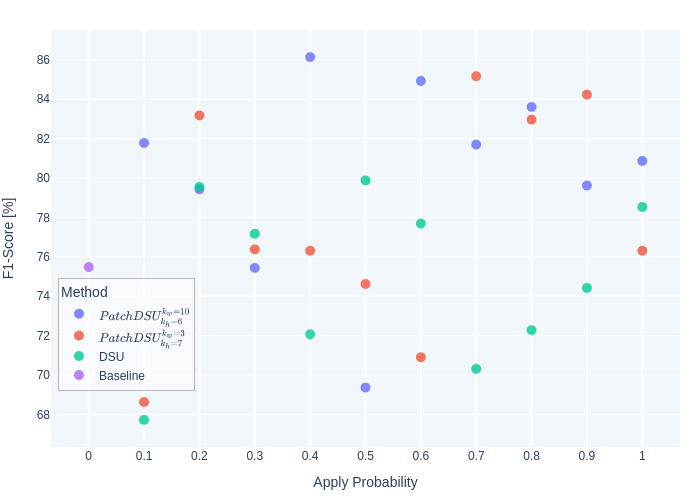}}\par
    \caption{Performance of models trained on TED-LIUM and tested on GSC and Librispeech keywords test splits (test-set without the Unknown class).}
    \label{fig:ted_ood}
\end{figure}

Lastly, when evaluating the test set of Google Speech Commands, the model trained with Freq-MixStyle outperformed other methods by a large margin. This behavior is unexpected, as the model demonstrated a lower F1-score than all other methods, including the baseline, as illustrated in Table \ref{tab:cleanacc}. Furthermore, across OODG scenarios from TED-LIUM to Librispeech and vice-versa, and from Google Speech Commands on TED-LIUM, the model underperformed compared to the baseline model. Analyzing the overall trend on this out-of-distribution test-set in Figure \ref{fig:ted_ood_on_gcom}, reveals that using higher $p$ values for {\method} and {\methodbase} models was preferable. Suspecting that the gap between Freq-MixStyle model and the other methods stems from the perturbation strength, we compared it to the best-performing model on this test set, {\method}$(k_h\!=\!6, k_w\!=\!10)$. Nonetheless, while the margin between models decreased, the Freq-MixStyle model performed best. This should be inspected in future work. Specifically, whether frequency-focused patches splits, could better account for specific out-of-distribution cases.

\section{Discussion}
In this work, we propose a method to improve out-of-domain generalization for keyword spotting. Particularly, we applied {\methodbase} on the task of keyword spotting and introduced an extension called {\method}, which operates on patches instead of the entire inputs. We evaluate the performance of the different methods on the signal with and without noise augmentation, and in out-of-domain scenarios. 

Our experiments show that {\method} generally enhances in-domain generalization compared to the other tested approaches, particularly in low-resource or imbalanced datasets, showcasing improved accuracy on the standard signal and the time-shifted signal as well as robustness to MUSAN noises. However, when considering white Gaussian noise (WGN) as well as MUSAN noises on Google Speech Commands, {\methodbase} should be considered. 

Furthermore, on out-of-domain generalization (OODG), we showed several cases where {\method} models outperformed other methods. In the other instances, except for OODG from TED-LIUM to Google Speech Commands, we either demonstrated additional {\method} models compensating for the performance gap or that the margin between the best model and {\method} models was not favorable compared to the generalization gap that the top model had with the proposed method. 

In addition, we showed that our initial hypothesis about the distribution variability among these datasets is consistent with the results. Generalizing from Librispeech to the other datasets and vice-versa, is better than generalization from Google Speech Commands to TED-LIUM and vice versa. For practical use cases, there is potential in exploring the leveraging of conditional decision-making, where models with a specific $p$ value will be applied based on the identification of a distribution gap.

While no method exceeds other approaches across all test cases,
which requires further investigation to develop strategies for consistent OODG; it should not overshadow the promising results of {\method}. Importantly, {\method} offers a single model that demonstrates more consistent performance across various generalization and out-of-domain generalization scenarios compared to the other approaches evaluated, where in most cases, {\method} performed better than the alternatives.


It is worth mentioning that, while in the original work of {\methodbase}, different $p$-values achieved similar results, in the speech domain, both {\method} and {\methodbase} require calibration of the $p$ values and consideration of the properties of the testing domain. Further research is needed to refine the calibration process of {\method} and potentially explore hybrid solutions to reconcile the performance disparities between {\method}$(k_h\!=\!7, k_w\!=\!3)$ and {\method}$(k_h\!=\!10, k_w\!=\!6)$, as they demonstrated varying strengths across different experiments.  
Future research could overcome this by exploring an extension to our approach, such as enhancing the flexibility of patches rather than maintaining fixed sizes throughout the layers. Moreover, further efforts could be made to bridge the observed gap between generalization and out-of-domain generalization, as our experiments showed in specific scenarios. Additional investigation could focus on datasets that align closely with our assumption of data sparsity. This, in turn, could provide valuable insights for further advancements in the field. 

\bibliographystyle{IEEEtran}
\bibliography{mybib}

\begin{thebibliography}{10}
\providecommand{\url}[1]{#1}
\csname url@samestyle\endcsname
\providecommand{\newblock}{\relax}
\providecommand{\bibinfo}[2]{#2}
\providecommand{\BIBentrySTDinterwordspacing}{\spaceskip=0pt\relax}
\providecommand{\BIBentryALTinterwordstretchfactor}{4}
\providecommand{\BIBentryALTinterwordspacing}{\spaceskip=\fontdimen2\font plus
\BIBentryALTinterwordstretchfactor\fontdimen3\font minus \fontdimen4\font\relax}
\providecommand{\BIBforeignlanguage}[2]{{%
\expandafter\ifx\csname l@#1\endcsname\relax
\typeout{** WARNING: IEEEtran.bst: No hyphenation pattern has been}%
\typeout{** loaded for the language `#1'. Using the pattern for}%
\typeout{** the default language instead.}%
\else
\language=\csname l@#1\endcsname
\fi
#2}}
\providecommand{\BIBdecl}{\relax}
\BIBdecl

\bibitem{ben2010theory}
S.~Ben-David, J.~Blitzer, K.~Crammer, A.~Kulesza, F.~Pereira, and J.~W. Vaughan, ``A theory of learning from different domains,'' \emph{Machine learning}, vol.~79, pp. 151--175, 2010.

\bibitem{vapnik1991principles}
V.~Vapnik, ``Principles of risk minimization for learning theory,'' \emph{Advances in neural information processing systems}, vol.~4, 1991.

\bibitem{wilson2020survey}
G.~Wilson and D.~J. Cook, ``A survey of unsupervised deep domain adaptation,'' \emph{ACM Transactions on Intelligent Systems and Technology (TIST)}, vol.~11, no.~5, pp. 1--46, 2020.

\bibitem{palaz2016jointly}
D.~Palaz, G.~Synnaeve, and R.~Collobert, ``Jointly learning to locate and classify words using convolutional networks.'' in \emph{Interspeech}, 2016, pp. 2741--2745.

\bibitem{segal2019speechyolo}
Y.~Segal, T.~S. Fuchs, and J.~Keshet, ``Speechyolo: Detection and localization of speech objects,'' \emph{Proc. Interspeech 2019}, pp. 4210--4214, 2019.

\bibitem{fuchs2021cnn0}
T.~S. Fuchs, Y.~Segal, and J.~Keshet, ``Cnn-based spoken term detection and localization without dynamic programming,'' in \emph{ICASSP 2021-2021 IEEE International Conference on Acoustics, Speech and Signal Processing (ICASSP)}.\hskip 1em plus 0.5em minus 0.4em\relax IEEE, 2021, pp. 6853--6857.

\bibitem{vsvec2017relevance}
J.~{\v{S}}vec, J.~V. Psutka, L.~{\v{S}}m{\'\i}dl, and J.~Trmal, ``A relevance score estimation for spoken term detection based on rnn-generated pronunciation embeddings,'' in \emph{Proc. Interspeech 2017}, 2017, pp. 2934--2938.

\bibitem{vsvec2022spoken}
J.~{\v{S}}vec, L.~{\v{S}}m{\'\i}dl, J.~V. Psutka, and A.~Pra{\v{z}}{\'a}k, ``Spoken term detection and relevance score estimation using dot-product of pronunciation embeddings,'' \emph{arXiv preprint arXiv:2210.11895}, 2022.

\bibitem{shin2022learning}
H.-K. Shin, H.~Han, D.~Kim, S.-W. Chung, and H.-G. Kang, ``Learning audio-text agreement for open-vocabulary keyword spotting,'' \emph{arXiv preprint arXiv:2206.15400}, 2022.

\bibitem{nishu2023matching}
K.~Nishu, M.~Cho, and D.~Naik, ``Matching latent encoding for audio-text based keyword spotting,'' \emph{arXiv preprint arXiv:2306.05245}, 2023.

\bibitem{chen2014small}
G.~Chen, C.~Parada, and G.~Heigold, ``Small-footprint keyword spotting using deep neural networks,'' in \emph{2014 IEEE international conference on acoustics, speech and signal processing (ICASSP)}.\hskip 1em plus 0.5em minus 0.4em\relax IEEE, 2014, pp. 4087--4091.

\bibitem{sainath2015convolutional}
T.~N. Sainath and C.~Parada, ``Convolutional neural networks for small-footprint keyword spotting.'' in \emph{Interspeech}, 2015, pp. 1478--1482.

\bibitem{zeng2019effective}
M.~Zeng and N.~Xiao, ``Effective combination of densenet and bilstm for keyword spotting,'' \emph{IEEE Access}, vol.~7, pp. 10\,767--10\,775, 2019.

\bibitem{tang2018deep}
R.~Tang and J.~Lin, ``Deep residual learning for small-footprint keyword spotting,'' in \emph{2018 IEEE International Conference on Acoustics, Speech and Signal Processing (ICASSP)}.\hskip 1em plus 0.5em minus 0.4em\relax IEEE, 2018, pp. 5484--5488.

\bibitem{coucke2019efficient}
A.~Coucke, M.~Chlieh, T.~Gisselbrecht, D.~Leroy, M.~Poumeyrol, and T.~Lavril, ``Efficient keyword spotting using dilated convolutions and gating,'' in \emph{ICASSP 2019-2019 IEEE International Conference on Acoustics, Speech and Signal Processing (ICASSP)}.\hskip 1em plus 0.5em minus 0.4em\relax IEEE, 2019, pp. 6351--6355.

\bibitem{choi2019temporal}
S.~Choi, S.~Seo, B.~Shin, H.~Byun, M.~Kersner, B.~Kim, D.~Kim, and S.~Ha, ``Temporal convolution for real-time keyword spotting on mobile devices,'' \emph{Interspeech 2019}, 2019.

\bibitem{li2020small}
X.~Li, X.~Wei, and X.~Qin, ``Small-footprint keyword spotting with multi-scale temporal convolution,'' \emph{Interspeech 2020}, 2020.

\bibitem{kim2021broadcasted}
B.~Kim, S.~Chang, J.~Lee, and D.~Sung, ``Broadcasted residual learning for efficient keyword spotting,'' \emph{arXiv preprint arXiv:2106.04140}, 2021.

\bibitem{sun2017unsupervised}
S.~Sun, B.~Zhang, L.~Xie, and Y.~Zhang, ``An unsupervised deep domain adaptation approach for robust speech recognition,'' \emph{Neurocomputing}, vol. 257, pp. 79--87, 2017.

\bibitem{su2024corpus}
H.~Su, T.-Y. Hu, H.~S. Koppula, R.~Vemulapalli, J.-H.~R. Chang, K.~Yang, G.~V. Mantena, and O.~Tuzel, ``Corpus synthesis for zero-shot asr domain adaptation using large language models,'' in \emph{ICASSP 2024-2024 IEEE International Conference on Acoustics, Speech and Signal Processing (ICASSP)}.\hskip 1em plus 0.5em minus 0.4em\relax IEEE, 2024, pp. 12\,326--12\,330.

\bibitem{wang2018unsupervised}
Q.~Wang, W.~Rao, S.~Sun, L.~Xie, E.~S. Chng, and H.~Li, ``Unsupervised domain adaptation via domain adversarial training for speaker recognition,'' in \emph{2018 IEEE international conference on acoustics, speech and signal processing (ICASSP)}.\hskip 1em plus 0.5em minus 0.4em\relax IEEE, 2018, pp. 4889--4893.

\bibitem{stafylakis2018zero}
T.~Stafylakis and G.~Tzimiropoulos, ``Zero-shot keyword spotting for visual speech recognition in-the-wild,'' in \emph{Proceedings of the European Conference on Computer Vision (ECCV)}, 2018, pp. 513--529.

\bibitem{lee2023phonmatchnet}
Y.-H. Lee and N.~Cho, ``Phonmatchnet: phoneme-guided zero-shot keyword spotting for user-defined keywords,'' \emph{arXiv preprint arXiv:2308.16511}, 2023.

\bibitem{mazumder2021few}
M.~Mazumder, C.~Banbury, J.~Meyer, P.~Warden, and V.~J. Reddi, ``Few-shot keyword spotting in any language,'' \emph{arXiv preprint arXiv:2104.01454}, 2021.

\bibitem{zhang2017mixup}
H.~Zhang, M.~Cisse, Y.~N. Dauphin, and D.~Lopez-Paz, ``mixup: Beyond empirical risk minimization,'' \emph{arXiv preprint arXiv:1710.09412}, 2017.

\bibitem{shankar2018generalizing}
S.~Shankar, V.~Piratla, S.~Chakrabarti, S.~Chaudhuri, P.~Jyothi, and S.~Sarawagi, ``Generalizing across domains via cross-gradient training,'' \emph{arXiv preprint arXiv:1804.10745}, 2018.

\bibitem{nam2018batch}
H.~Nam and H.-E. Kim, ``Batch-instance normalization for adaptively style-invariant neural networks,'' \emph{Advances in Neural Information Processing Systems}, vol.~31, 2018.

\bibitem{piratla2020efficient}
V.~Piratla, P.~Netrapalli, and S.~Sarawagi, ``Efficient domain generalization via common-specific low-rank decomposition,'' in \emph{International Conference on Machine Learning}.\hskip 1em plus 0.5em minus 0.4em\relax PMLR, 2020, pp. 7728--7738.

\bibitem{kim2022domain}
B.~Kim, S.~Yang, J.~Kim, H.~Park, J.~Lee, and S.~Chang, ``Domain generalization with relaxed instance frequency-wise normalization for multi-device acoustic scene classification,'' \emph{arXiv preprint arXiv:2206.12513}, 2022.

\bibitem{zhou2021domain}
K.~Zhou, Y.~Yang, Y.~Qiao, and T.~Xiang, ``Domain generalization with mixstyle,'' \emph{arXiv preprint arXiv:2104.02008}, 2021.

\bibitem{li2022uncertainty}
\BIBentryALTinterwordspacing
X.~Li, Y.~Dai, Y.~Ge, J.~Liu, Y.~Shan, and L.~DUAN, ``Uncertainty modeling for out-of-distribution generalization,'' in \emph{International Conference on Learning Representations}, 2022. [Online]. Available: \url{https://openreview.net/forum?id=6HN7LHyzGgC}
\BIBentrySTDinterwordspacing

\bibitem{warden2018speech}
P.~Warden, ``Speech commands: A dataset for limited-vocabulary speech recognition,'' \emph{arXiv preprint arXiv:1804.03209}, 2018.

\bibitem{panayotov2015librispeech}
V.~Panayotov, G.~Chen, D.~Povey, and S.~Khudanpur, ``Librispeech: an asr corpus based on public domain audio books,'' in \emph{2015 IEEE international conference on acoustics, speech and signal processing (ICASSP)}.\hskip 1em plus 0.5em minus 0.4em\relax IEEE, 2015, pp. 5206--5210.

\bibitem{hernandez2018ted}
F.~Hernandez, V.~Nguyen, S.~Ghannay, N.~Tomashenko, and Y.~Esteve, ``Ted-lium 3: Twice as much data and corpus repartition for experiments on speaker adaptation,'' in \emph{Speech and Computer: 20th International Conference, SPECOM 2018, Leipzig, Germany, September 18--22, 2018, Proceedings 20}.\hskip 1em plus 0.5em minus 0.4em\relax Springer, 2018, pp. 198--208.

\bibitem{shen2021closed}
Y.~Shen and B.~Zhou, ``Closed-form factorization of latent semantics in gans,'' in \emph{Proceedings of the IEEE/CVF conference on computer vision and pattern recognition}, 2021, pp. 1532--1540.

\bibitem{wang2019implicit}
Y.~Wang, X.~Pan, S.~Song, H.~Zhang, G.~Huang, and C.~Wu, ``Implicit semantic data augmentation for deep networks,'' \emph{Advances in Neural Information Processing Systems}, vol.~32, 2019.

\bibitem{kingma2013auto}
D.~P. Kingma and M.~Welling, ``Auto-encoding variational bayes,'' \emph{arXiv preprint arXiv:1312.6114}, 2013.

\bibitem{ioffe2015batch}
S.~Ioffe, ``Batch normalization: Accelerating deep network training by reducing internal covariate shift,'' \emph{arXiv preprint arXiv:1502.03167}, 2015.

\bibitem{ulyanov2016instance}
D.~Ulyanov, A.~Vedaldi, and V.~Lempitsky, ``Instance normalization: The missing ingredient for fast stylization,'' \emph{arXiv preprint arXiv:1607.08022}, 2016.

\bibitem{baevski2020wav2vec}
A.~Baevski, Y.~Zhou, A.~Mohamed, and M.~Auli, ``wav2vec 2.0: A framework for self-supervised learning of speech representations,'' \emph{Advances in neural information processing systems}, vol.~33, pp. 12\,449--12\,460, 2020.

\bibitem{pratap2023scaling}
V.~Pratap, A.~Tjandra, B.~Shi, P.~Tomasello, A.~Babu, S.~Kundu, A.~Elkahky, Z.~Ni, A.~Vyas, M.~Fazel-Zarandi \emph{et~al.}, ``Scaling speech technology to 1,000+ languages,'' \emph{arXiv preprint arXiv:2305.13516}, 2023.

\bibitem{snyder2015musan}
D.~Snyder, G.~Chen, and D.~Povey, ``Musan: A music, speech, and noise corpus,'' \emph{arXiv preprint arXiv:1510.08484}, 2015.

\bibitem{he2015deep}
K.~He, X.~Zhang, S.~Ren, and J.~Sun, ``Deep residual learning for image recognition. arxiv e-prints,'' \emph{arXiv preprint arXiv:1512.03385}, vol.~10, 2015.

\bibitem{agarap2018deep}
A.~Agarap, ``Deep learning using rectified linear units (relu),'' \emph{arXiv preprint arXiv:1803.08375}, 2018.

\bibitem{loshchilov2016sgdr}
I.~Loshchilov and F.~Hutter, ``Sgdr: Stochastic gradient descent with warm restarts,'' \emph{arXiv preprint arXiv:1608.03983}, 2016.

\bibitem{goyal2017accurate}
P.~Goyal, P.~Doll{\'a}r, R.~Girshick, P.~Noordhuis, L.~Wesolowski, A.~Kyrola, A.~Tulloch, Y.~Jia, and K.~He, ``Accurate, large minibatch sgd: Training imagenet in 1 hour,'' \emph{arXiv preprint arXiv:1706.02677}, 2017.

\end{thebibliography}
\section{Appendix}

\end{document}